\documentclass[intlimits,twoside,a4paper]{article}
\usepackage{bbm}
\usepackage{mathrsfs}
\usepackage{epsfig,psfrag}
\usepackage[final]{pdfpages}
\usepackage{amsmath,amsfonts,amssymb}

\usepackage{graphicx}
\usepackage[cp1251]{inputenc}

\usepackage[eqsecnum]{cmpj3}

\def\be{\begin{equation}}
\def\ee{\end{equation}}
\def\bea{\begin{eqnarray}}
\def\eea{\end{eqnarray}}

\newcommand{\pd}{\partial}

\def\smat#1{\left(\begin{matrix}#1\end{matrix}\right)}

\issue{2017}{20}{2}{23701}
\doinumber{10.5488/CMP.20.23701}

\title[Ground state properties of the bond alternating
spin-$\frac{1}{2}$ anisotropic Heisenberg chain]%
{Ground state properties of the bond alternating
spin-$\frac{1}{2}$ anisotropic Heisenberg chain%
}
\author[S. Paul, A.K. Ghosh]{S. Paul\refaddr{label1},
        A.K. Ghosh\refaddr{label2}}
\addresses{
\addr{label1} Department of Physics, Scottish Church College,
Urquhart Square, Kolkata 700006, India
\addr{label2} Department of Physics, Jadavpur University,
188 Raja Subodh Chandra Mallik Road, Kolkata 700032, India
}

\date{Received October 16, 2016, in final form December 17, 2016}

\begin{document}

\maketitle

\begin{abstract}
Ground state properties, dispersion relations and scaling behaviour of spin gap
of a bond alternating  spin-$\frac{1}{2}$ anisotropic Heisenberg
 chain  have been studied where the  exchange
interactions on alternate bonds are ferromagnetic (FM)
and antiferromagnetic (AFM) in two separate cases.
The resulting models separately represent nearest neighbour (NN) AFM-AFM and AFM-FM bond
alternating chains.
Ground state energy has been estimated analytically
by using both bond operator and Jordan-Wigner representations
and numerically by using exact diagonalization.
Dispersion relations, spin gap and several ground state orders have been obtained.
Dimer order and string orders are found to coexist in the ground state.
Spin gap is found to develop as soon as the non-uniformity in
alternating bond strength is introduced in the AFM-AFM chain
which further remains non-zero for the AFM-FM chain.
This spin gap along with the string orders attribute to the Haldane phase. The
Haldane phase is found to exist in most of the anisotropic region
similar to the isotropic point.

\pacs 75.10.Jm, 75.10.Pq, 75.50.Ee, 75.40.Mg, 75.30.Gw
\end{abstract}

\section{Introduction}
The spin-$\frac{1}{2}$ Heisenberg chains with an energy gap (spin gap)
just above the ground state attract immense interest since they give rise to
many exotic properties in the ground state.
The isotropic AFM and FM
spin-$\frac{1}{2}$ Heisenberg chains
are exactly solvable by using the Bethe-Ansatz technique
both in the presence and in the absence of uniform magnetic
field in which energy
spectrum is gapless below a critical field \cite{Griffiths}.
On the other hand, according to Haldane's conjecture \cite{Haldane},
AFM Heisenberg chain with integer spin values has a finite spin gap
between non-magnetic ground state and the lowest excited state
which is known as the Haldane gap. The Haldane phase can be characterized by
the finite value of string order parameter \cite{Rommelse,Tasaki}.
The existence of this spin gap can be explained from the incongruousness of this
system with the Lieb-Schultz-Mattis (LSM) theorem \cite{LSM}.
According to the modified version of
LSM theorem extended by Affleck and Lieb \cite{LSMA},
the SU(2) invariant AFM chains
with half-integer spins per unit cell either have gapless excitations
or degenerate ground states in the thermodynamic limit,
$N\!\rightarrow\! \infty$.
Finally, it has been extended to more than one dimension and shown
to be valid for short range interactions with global U(1)
symmetry and half-integer spin per unit cell \cite{Hastings}.

A spin gap in AFM bond alternating spin-$\frac{1}{2}$ Heisenberg chain was first
predicted theoretically in 1962 by Bulaevskii \cite{Bulaevskii}.
The nature of triplet excitations at finite temperatures \cite{Harris}
and multimagnon excitations \cite{Southern} in bond alternating
chain has been studied.
Hidden $Z_2\times Z_2$ symmetry breaking along with the Haldane phase
is found by Kohmoto \cite{Kohmoto}. Magnetization process in anisotropic
bond alternating chain has been investigated by Totsuka \cite{Totsuka}.
In the bond alternating spin-$\frac{1}{2}$ Heisenberg chains,
the full translational symmetry of
the lattice is lost since a unit cell contains two lattice sites.
These two spin-$\frac{1}{2}$\linebreak degrees of freedom
combine to form either total spin 0 or 1. This situation does not
comply with the LSM theorem though the system
has global U(1) symmetry.
From this point of view, a gap in the
spin excitation may appear in the bond alternating chain.
The existence of Haldane gap is found
in an exactly solvable AFM one-dimensional
bilinear-biquadratic spin-1 model where the ground state
has a valence bond solid structure in which each integer spin value is
described as a symmetric combination of two half-integer spins forming
a singlet state within each pair of adjacent sites \cite{AKLT1,AKLT2}.
In 1992, Hida pointed out that isotropic $S=\frac{1}{2}$ Heisenberg chain
with alternating AFM and FM couplings can be mapped
onto the isotropic $S=1$ AFM Heisenberg chain when the FM couplings tend to
infinity \cite{Hida}. Therefore, the existence of Haldane phase
can be justified in the  $S=\frac{1}{2}$ Heisenberg chains
with bond alternation. A transition from Haldane phase to
gapless phase has been predicted in the presence of magnetic field \cite{Sakai}.
A number of compounds are discovered whose properties can be explained by
invoking either AFM-AFM or AFM-FM types of bond alternating chains.
For examples,
the compounds like CuGeO$_3$ \cite{Hase},
tetrathiafulvalene (TTF) with bisdithiolene metal complexes \cite{Jecobs},
TTFCuBDT \cite{Cross}, MEM-(TCNQ)$_3$ \cite{Huizinga} and many others
which show spin-Peierls transitions belong to AFM-AFM  bond alternating
class. On the other hand,
zinc-verdazyl complex \cite{Yamaguchi}, $\alpha$-CuNb$_2$O$_6$ \cite{Watanabe,Kodama},
Na$_3$Cu$_2$SbO$_6$ \cite{Miura}, (CH$_3$)$_2$CHNH$_3$CuCl$_3$ \cite{Manaka},
and DMACuCl$_3$ \cite{Stone}
belongs to the $S=\frac{1}{2}$ AFM-FM bond alternating
class.

In this work, anisotropic bond alternating
 $S=\frac{1}{2}$ Heisenberg chains
with alternating  AFM-AFM and AFM-FM couplings
have been studied separately where the ground state energy, dispersion relations,
ground state orders and the
magnitude of spin gap have been obtained for the
entire range of anisotropic parameters. Two different theoretical
approaches, say, bond operator and Jordan-Wigner representations
are employed in which the spin model is expressed in terms of
bosonic and fermionic operators, respectively. Mean-field analysis on these two approaches
gives rise to accurate results of this model in two different regimes. Ground state energy,
dispersion relations, dimer order and spin gap are obtained by using the
bond operator formalism. All those properties in addition to
string orders have been separately estimated by using exact diagonalization method.
Coexistence of dimer and string order parameters has been
found. The existence of the spin gap along with the string orders found
in most of the anisotropic region
attributes to the Haldane phase. We should like to report that this observation
is similar to that found at the isotropic point of these models as predicted before \cite{Hida}.

The bond alternating spin model is defined by the Hamiltonian
\bea
H=\sum_{i=1}^{N/2}\left[J_1\left(
S_{2i-1}^xS_{2i}^x+S_{2i-1}^yS_{2i}^y+\Delta S_{2i-1}^zS_{2i}^z\right)+
J_2\left(S_{2i}^xS_{2i+1}^x+S_{2i}^yS_{2i+1}^y+\Delta S_{2i}^zS_{2i+1}^z\right) \right].
\label{ham}
\eea
$N$ is the total number of spins which is even.
 The model has the global U(1) symmetry
since the $z$-component of the total spin,
$S^z_{\rm T}$, is a good quantum number. The $J_1$ bond is
always AFM but the $J_2$ bond is considered both AFM and FM, such that
$-1.0<\frac{J_2}{J_1}<1.0$. $\Delta$ is the anisotropic parameter.
 For $J_1=J_2$, the system remains gapless throughout the anisotropic
regime $0\leqslant \Delta\leqslant 1$, while the spin gap
opens up when $J_1 \neq J_2$.

Section~\ref{four-spin} contains the results obtained for a four-spin
bond alternating plaquette.
In sections~\ref{B-O} and~\ref{J-W}, investigations based on
bond operator and Jordan-Wigner representations
are described, respectively. The spin model is studied numerically by
using Lanczos exact diagonalization technique where ground state energy and
spin gap are obtained and reported in section~\ref{E-D}.
Values of several ground state orders
have been estimated and described in section~\ref{ground-properties}.
Section~\ref{conclusion} contains a discussion of the
results obtained.
\section{Four-site bond alternating anisotropic Heisenberg plaquette}
\label{four-spin}
\begin{figure}[!t]
\begin{center}
\psfrag{J2byJ1}{$J_2/J_1$}
\psfrag{do}{$\mathcal O_{\rm D}$}
\psfrag{a}{$\Delta=0.0$}
\psfrag{b}{$\Delta=0.3$}
\psfrag{c}{$\Delta=0.7$}
\psfrag{d}{$\Delta=1.0$}
\includegraphics[scale=1.1]{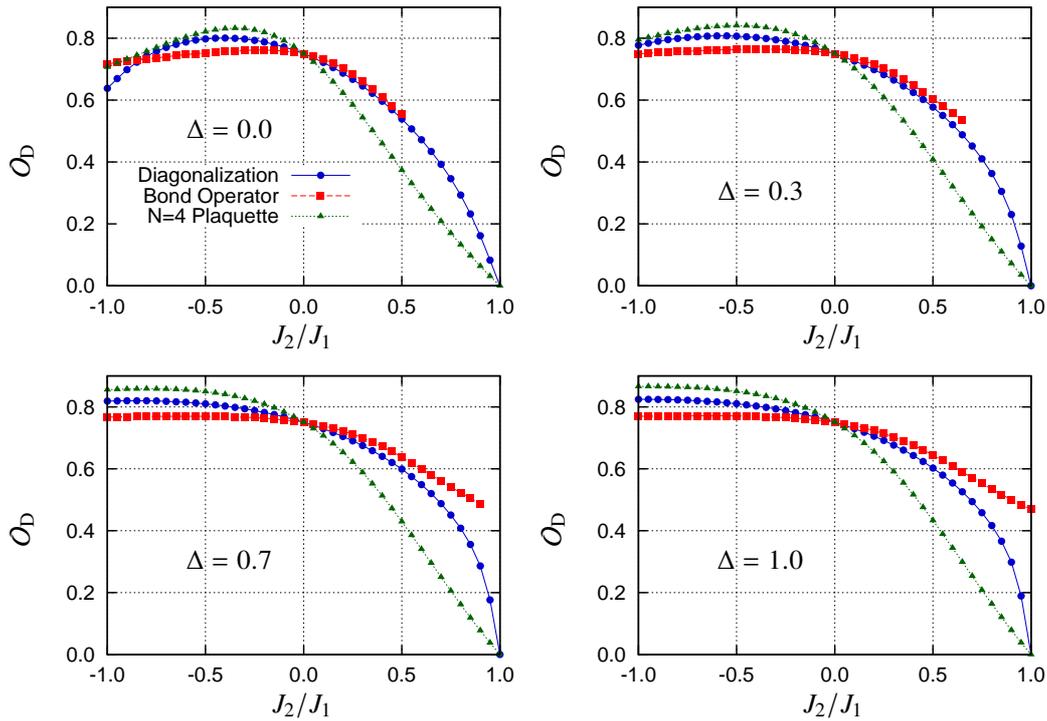}
\caption{(Color online)
Plots of dimer order as a function of $J_2/J_1$ for
$\Delta=0.0$, $0.3$, $0.7$, $1.0$.
Blue (solid line, circle), red (dashed line, square) and green (dotted line, triangle) correspond
to the exact diagonalization, bond operator and $N=4$ plaquette results, respectively.
Bond operator lines terminate at those points where the convergence of
self-consistent equations is not attained.}
\label{do}
\end{center}
\end{figure}

Before the beginning of an extensive  many-particle formalism, let us explain
the results of a four-spin ($N=4$) bond alternating
$S=\frac{1}{2}$ anisotropic Heisenberg plaquette.
Here, the stronger AFM bonds~($J_1$) are assumed between
the site-pairs $(1,2)$ and $(3,4)$, while the FM or weaker AFM bonds ($J_2$) are
acting between the site-pairs $(2,3)$ and $(4,1)$.
The Hamiltonian has been diagonalised in different $S^z_{\rm T}$
sectors for obtaining analytic expressions of eigenvalues and
eigenfunctions. Eigenvalues ($e_i$) are displayed in table~\ref{eigenvalues}.
Ground state lies in $S^z_{\rm T}=0$ sector having energy $e_0$.
The ground state wave function is given by
$\Psi_0=\frac{1}{\sqrt{1+X^2+Y^2}}(\psi_1+X\psi_2+Y\psi_3)$,
where $\psi_1=\frac{1}{\sqrt 2}(\uparrow\uparrow\downarrow\downarrow+\downarrow\downarrow\uparrow\uparrow)$,
 $\psi_2=\frac{1}{\sqrt 2}(\uparrow\downarrow\uparrow\downarrow+\downarrow\uparrow\downarrow\uparrow)$,
$\psi_3=\frac{1}{\sqrt 2}(\uparrow\downarrow\downarrow\uparrow+\downarrow\uparrow\uparrow\downarrow)$,
$X=\frac{e_0-e_4}{J_2}$ and $Y=X\frac{J_1}{e_0+e_4}$.  Dimer order parameter is defined by
the ground state expectation value \cite{White},
\[\mathcal{O}_{\rm D}=\langle \Psi_0| \vec S_i\cdot \vec S_{i+1} -
\vec S_{i+1} \cdot \vec S_{i+2}|\Psi_0\rangle
=\frac{1-2X-Y^2+2XY}{2(1+X^2+Y^2)}\,.\]
When $J_1=J_2$, the ground state energy, $e_0=-\frac{J_1}{2}(\Delta+\sqrt{8+\Delta^2})$,
and $\mathcal{O}_{\rm D}=0$ since $Y=1$.
On the FM region, when $J_2=-J_1$, $e_0=-J_1\sqrt{2+\Delta^2}$,
$\mathcal{O}_{\rm D}=(1-2x-x^2/2)/[2(1+x^2/2)]$,
where $x=(\Delta+\sqrt{2+\Delta^2})$.
When $\Delta=1$, ground state energy,
$e_0=-\frac{1}{2}[J_1+J_2+\sqrt{(J_1+J_2)^2+3(J_1-J_2)^2}]$
and ground state wave function, $\Psi_0=\frac{2}{\sqrt{1+c^2+d^2}}\left(c[14][32]+d[12][34]\right)$, where,
$c=\frac{J_2}{e_0-e_7}$,  $d=\frac{J_1}{e_0+e_7}$ and the singlet, [$ij$] is defined as
\[
[ij]=\frac{1}{\sqrt{2}}
\begin{array}{ccccc}
&&&&\\
(\uparrow & \downarrow & - & \downarrow & \uparrow).\\
\,\,i&j& &i&j \,\,\,
\end{array}
\]
However, when $\Delta=0$, ground state energy, $e_0=-\sqrt{J_1^2+J_2^2}$, and this is the same for both AFM and FM $J_2$
and ground state,  $\Psi_0=\frac{2e_0^0}{\sqrt{e_0^{0\;2}+J_1^2+J_2^2}}
(\frac{J_2}{e_0^0}[14][32]+\frac{J_1}{e_0^0}[12][34]-\frac{e_0^0+J_1+J_2}{\sqrt{2}e_0^0}\psi_2)$.
Variations of $\mathcal{O}_{\rm D}$ are shown in figure~\ref{do} in the green
dotted lines and triangles. $\mathcal{O}_{\rm D}$ vanishes
exactly over the line $J_2/J_1=1.0$ and otherwise non-zero. It is observed that
$\mathcal{O}_{\rm D}$ calculated in this four-spin bond alternating plaquette,
captures the true many-particle results closely.
\begin{table}[!t]
\caption{\label{eigenvalues} Eigenvalues in all $S^z_{\rm T}$ subspaces, where
$a=\frac{2}{3}[(\Delta^2+3)(J_1^2+J_2^2)-J_1J_2\Delta^2]^{1/2}$,
$b=-\frac{J_1+J_2}{6}\,\Delta$ and $\phi=\arccos[3q/(ap)]$,
$q=e_4(e_3e_4+J_1^2-J_2^2)-\frac{e_3}{27}[2e_3^2+9(e_4^2+J_1^2+J_2^2)]$ and
$p=-\frac{1}{3}[(\Delta^2+3)(J_1^2+J_2^2)-J_1J_2\Delta^2]$. }
\vspace{2ex}
\begin{center}
\begin{tabular}{|c|l|c|l| }
 \hline\hline
 $S^z_{\rm T}$  & Eigenvalues &  $S^z_{\rm T}$  & Eigenvalues \\  \hline\hline
  & $e_5=-\frac{J_1-J_2}{2}\,\Delta $ &   & $e_9=\frac{J_1+J_2}{2} $\\ 
  & $e_4=\frac{J_1-J_2}{2}\,\Delta $ & $1, -1$ & $e_8=\frac{-J_1+J_2}{2} $ \\ 
0 & $e_3=-\frac{J_1+J_2}{2}\,\Delta $&  & $e_7=\frac{J_1-J_2}{2} $ \\ 
 & $e_2=a\cos{\frac{\phi}{3}}+b $&   & $e_6=-\frac{J_1+J_2}{2} $ \\ \cline{3-4}
 & $e_1=a\cos{\frac{\phi+4\piup}{3}}+b $&$2, -2$ & $e_{10}= \frac{J_1+J_2}{2}\,\Delta $ \\
 & $e_0=a\cos{\frac{\phi+2 \piup}{3}}+b $& & \\
 \hline\hline
\end{tabular}
\end{center}
\end{table}

\section{Bond operator representation}
\label{B-O}
\begin{figure}[!b]
\begin{center}
\psfrag{J2byJ1}{$J_2/J_1$}
\psfrag{genergy}{$E_{\rm G}/J_1$}
\psfrag{a}{$\Delta=0.0$}
\psfrag{b}{$\Delta=0.3$}
\psfrag{c}{$\Delta=0.7$}
\psfrag{d}{$\Delta=1.0$}
\includegraphics[scale=1.05]{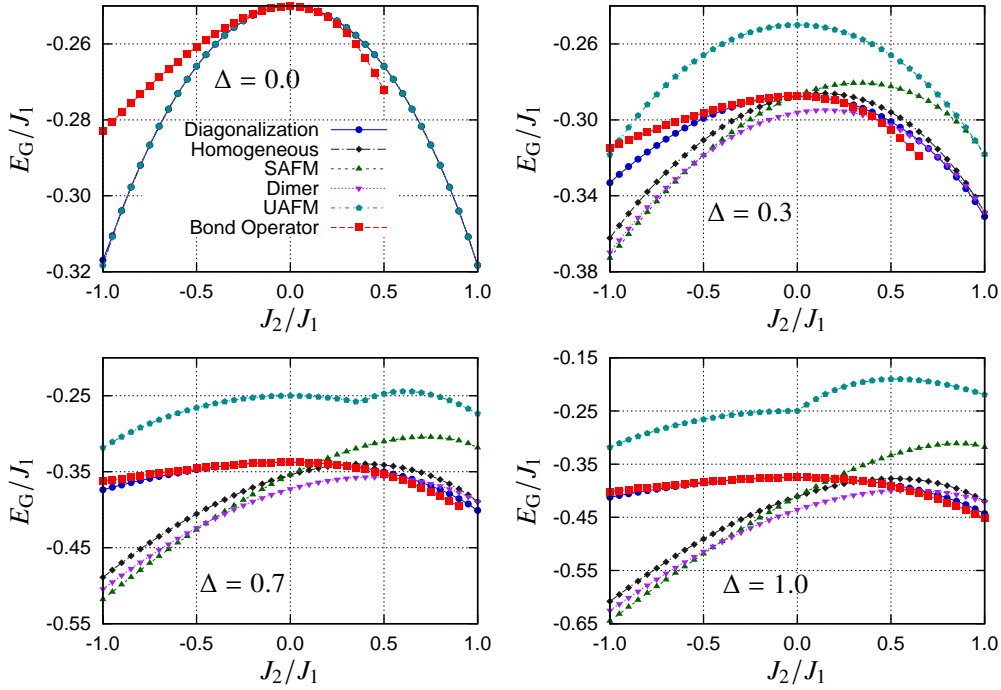}
\caption{\label{genergy}{(Color online)
Plots of ground state energy per site as a function of $J_2/J_1$ for
$\Delta=0.0$, $0.3$, $0.7$, $1.0$.
Blue (solid line, circle) corresponds to the exact diagonalization data.
Red (dashed line, square) corresponds to the bond operator result.
Different Jordan-Wigner based mean-field results:
UAFM (dark-cyan, dashed-dot line, pentagon),
homogeneous (black, dashed-dot line, diamond),
SAFM (green, dashed line, triangle) and
dimer (purple, dotted line, inverted triangle).
Bond operator lines terminate at those points where the convergence of
self-consistent equations is not attained. }}
\end{center}
\end{figure}

In the bond operator formalism \cite{sachdev}, two spin-$\frac{1}{2}$
operators, say, $\vec S_l$ and  $\vec S_r$ around every
AFM bond having exchange strength $J_1$ are expressed
in terms of a singlet state $|s\rangle$ and three triplet
states $|t_x\rangle$, $|t_y\rangle$, and $|t_z\rangle$ around the same bond.
The singlet ($s^\dagger$) and triplet ($t_\alpha^\dagger$, $\alpha=x,y,z$)
operators which create these states out of the vacuum state $|0\rangle$,
are
\begin{align}
|s\rangle &= s^\dagger|0\rangle =\frac{1}{\sqrt 2}
\left(|\uparrow \downarrow \rangle -|\downarrow \uparrow \rangle \right),\nonumber\\
|t_x\rangle &= t_x^\dagger|0\rangle =-\frac{1}{\sqrt 2}
\left(|\uparrow \uparrow  \rangle -|\downarrow\downarrow  \rangle \right),\nonumber
\end{align}
\begin{align}
|t_y\rangle &= t_y^\dagger|0\rangle =\frac{\ri}{\sqrt 2}
\left(|\uparrow \uparrow  \rangle +|\downarrow\downarrow  \rangle \right),\nonumber\\
|t_z\rangle &= t_z^\dagger|0\rangle =\frac{1}{\sqrt 2}
\left(|\uparrow \downarrow \rangle +|\downarrow \uparrow \rangle \right).\nonumber
\end{align}
Only the singlet state changes sign upon interchanging the two spins in each bond.
The components of spin operators, $\vec S_l$ and  $\vec S_r$ can be expressed in terms of
these singlet and triplet operators as
\begin{align}
S_l^\alpha&=\frac{1}{2}\left( s^\dagger \,t_\alpha+ t_\alpha^\dagger\, s-
\ri\,\epsilon_{\alpha\beta\gamma}\,t_\beta^\dagger\, t_\gamma\right), \nonumber\\
S_r^\alpha&=\frac{1}{2}\left(- s^\dagger \,t_\alpha- t_\alpha^\dagger\, s-
\ri\,\epsilon_{\alpha\beta\gamma}\,t_\beta^\dagger \,t_\gamma\right),
\label{spincompo}
\end{align}
where $\alpha$, $\beta$ and $\gamma$ represent the $x$, $y$ and $z$ components
and the Levi-Civit\'a symbol, $\epsilon_{\alpha\beta\gamma}$ represents the
totally anti-symmetric tensor. Summation over the repeated $\alpha$, $\beta$ and $\gamma$
indices is henceforth assumed except stated otherwise.
By considering the bosonic commutation relations, like
\[
[s,\,s^\dagger]=1,\qquad [t_\alpha,\,t_\beta^\dagger]=\delta_{\alpha\beta}\,,
\qquad [s,\,t_\alpha^\dagger]=0,
\]
on a particular bond,
one can reproduce the $S=\frac{1}{2}$, SU(2) commutation relations on a specific site,
$[S^\alpha,\,S^\beta]=\ri\,\epsilon_{\alpha\beta\gamma}\,S^\gamma$. Similarly, by imposing
the constraint, or the completeness relation,
\be
s^\dagger\,s+t^\dagger_\alpha\,t_\alpha=1,
\label{constraint}
\ee
on each bond, the value of spin in each site, say, $S_l^2=S_r^2=\frac{3}{4}$ is retained.
Likewise, the anisotropic AFM bond can be expressed as
\[
S_{2i-1}^xS_{2i}^x+S_{2i-1}^yS_{2i}^y+\Delta S_{2i-1}^zS_{2i}^z=E_s\,s_j^\dagger\,s_j
+E_t^z \,t^\dagger_{j\,z}\,t_{j\,z}+E_t^{\alpha}\,t^\dagger_{j\,\alpha}\,t_{j\,\alpha}\,,
\]
 where $j$ represents the bond between the adjacent sites $2i-1$ and $2i$,
$\alpha=x,y$, $E_s=-(\frac{\Delta}{4}+\frac{1}{2})$ is the singlet while
$E_t^z=(-\frac{\Delta}{4}+\frac{1}{2})$ along with the doubly degenerate
$E_t^{\alpha}=\frac{\Delta}{4}$ are the triplet eigenvalues of the anisotropic bond.
Substituting the operator representation of spins defined in equation~(\ref{spincompo}) into the
bond alternating Hamiltonian of equation~(\ref{ham}) we have the form:
\begin{align}
H&=H_0+H_1+H_2\,, \nonumber\\
H_0&=\sum_j\left[J_1\left(E_s\,s_j^\dagger\,s_j+E_t^z \,t^\dagger_{j\,z}\,t_{j\,z}
+E_t^{\alpha}\,t^\dagger_{j\,\alpha}\,t_{j\,\alpha} \right)
-\mu\left(s_j^\dagger\,s_j+t^\dagger_{j\,z}\,t_{j\,z}+
t^\dagger_{j\,\alpha}\,t_{j\,\alpha}-1  \right)\right],\nonumber \\
H_1&=-\frac{J_2}{4}\sum_j\left(s_j^\dagger s_{j+1}t_{j\,\alpha}t_{j+1\,\alpha}^\dagger+
 s_j s_{j+1}t_{j\,\alpha}^\dagger t_{j+1\,\alpha}^\dagger
+t_{j\,z}^\dagger t_{j+1\,z}^\dagger t_{j\,\alpha} t_{j+1\,\alpha}
-t_{j\,z}^\dagger t_{j+1\,z} t_{j\,\alpha} t_{j+1\,\alpha}^\dagger+\text{h.c.} \right),\nonumber \\
H_2&=-\frac{J_2\Delta}{4}\sum_j\left(s_j^\dagger s_{j+1}t_{j\,z}t_{j+1\,z}^\dagger+
 s_j s_{j+1}t_{j\,z}^\dagger t_{j+1\,z}^\dagger+t_{j\,x}^\dagger t_{j+1\,x}^\dagger t_{j\,y} t_{j+1\,y}
- t_{j\,x}^\dagger t_{j+1\,x} t_{j\,y} t_{j+1\,y}^\dagger+\text{h.c.}\right),
\label{bondham}
\end{align}
where the summation $j$ runs over $N/2$ number of bonds.
The portion of the Hamiltonian containing triple-$t$ operators vanishes
due to reflection symmetry \cite{sachdev}.
Exploiting the translational invariance of the model, a site-independent parameter
$\mu$  is introduced
to take the constraint, equation~(\ref{constraint}) into care. Here, condensation of
singlet boson is imposed, which means $\langle s_j\rangle=\bar s$.
Parts of the Hamiltonian, $H_1$ and $H_2$, those containing quartic $t$  operators are treated
by using mean-field decoupling scheme. Four mean-field parameters (real) are
\be
P_z=\langle t_{j\,z}^\dagger t_{j+1\,z} \rangle, \qquad
P_\alpha=\langle t_{j\,\alpha}^\dagger t_{j+1\,\alpha} \rangle, \qquad
Q_z=\langle t_{j\,z} t_{j+1\,z} \rangle \qquad \text{and}\qquad
Q_\alpha=\langle t_{j\,\alpha} t_{j+1\,\alpha} \rangle.
\ee
Summation convention over $\alpha$ while defining $P_\alpha$ and
$Q_\alpha$ is not applied.
By performing Fourier transform of the operators
$t_{j}=\sqrt{\frac{2}{N}}\sum_k t_k\,\re^{\ri kja}$, where $a$ is the
lattice constant, the approximated Hamiltonian can be written as
\be
H_{\text M}=E_0+\sum_k\left[ \Lambda_{k\,z}t_{k\,z}^\dagger t_{k\,z}+
\Lambda_{k\,\alpha}t_{k\,\alpha}^\dagger t_{k\,\alpha}+\Delta_{k\,z} \left( t_{k\,z} t_{-k\,z}+\text{h.c.}\right)
+\Delta_{k\,\alpha} \left( t_{k\,\alpha} t_{-k\,\alpha}+\text{h.c.}\right) \right],
\ee
where
\begin{align}
E_0&=\frac{N}{2}\left\{ \mu-\left[ J_1 \left( \frac{\Delta}{4}+\frac{1}{2}\right)+\mu \right]{\bar s} ^2
+J_2\left[ Q_\alpha Q_z-P_\alpha P_z+\frac{\Delta}{2}\left( Q_\alpha^2-P_\alpha^2\right)\right] \right\},\nonumber \\
\Lambda_{k\,z}&=J_1 \left( -\frac{\Delta}{4}+\frac{1}{2}\right)-\mu
-\frac{J_2}{2}\left(\Delta \,{\bar s} ^2 -2 P_\alpha \right)\cos(ka),\nonumber \\
\Lambda_{k\,\alpha}&=J_1 \frac{\Delta}{4}-\mu
-\frac{J_2}{2}\left({\bar s} ^2 - P_z -\Delta \, P_\alpha \right)\cos(ka),\nonumber \\
\Delta_{k\,z}&=-\frac{J_2}{4}\left(\Delta \,{\bar s} ^2 +2 Q_\alpha \right)\cos(ka),\qquad
\Delta_{k\,\alpha}=-\frac{J_2}{4}\left({\bar s} ^2 + Q_z +\Delta \, Q_\alpha \right)\cos(ka).\nonumber
\end{align}
The Hamiltonian, $H_{\text M}$ can be easily diagonalized by introducing the four-component vector
$\Psi_k=(t_{k\,z}^\dagger\; t_{k\,\alpha}^\dagger\; t_{-k\,z} \;  t_{-k\,\alpha})$. Thus,
 $H_{\text M}$ can be expressed as
\begin{align}
 H_{\text M}&=E_0-\frac{1}{2}\sum_k\left(\Lambda_{k\,z}+2\,\Lambda_{k\,\alpha}\right)
+ \sum_k \Psi_k^\dagger\, H_k \,\Psi_k\, ,\nonumber \\
H_k&=\smat{A_k&B_k\\B_k&A_k},\qquad  A_k=\frac{1}{2}\smat{\Lambda_{k\,z}&0\\0&\Lambda_{k\,\alpha} },\qquad
B_k=\smat{\Delta_{k\,z}&0\\0&\Delta_{k\,\alpha} }. \nonumber
\end{align}
In terms of bosonic operators, Bogoliubov transformation means diagonalization of
the matrix $I_BH_k$, where
\[
I_B=\smat{I&0\\0&-I},\qquad {\rm and}\quad I=\smat{1&0\\0&1}.
\]
The positive eigenvalues of the matrix, $I_BH_k$ are $\frac{1}{2}\omega_{k\,z}$ and
$\frac{1}{2}\omega_{k\,\alpha}$, where
$\omega_{k\,z}=\sqrt{\Lambda_{k\,z}^2-4\Delta_{k\,z}^2}$ and
$\omega_{k\,\alpha}=\sqrt{\Lambda_{k\,\alpha}^2-4\Delta_{k\,\alpha}^2}$.
In terms of a new four-component vector
$\Phi_k^\dagger =(\gamma_{k\,z}^\dagger\; \gamma_{k\,\alpha}^\dagger\;
\gamma_{-k\,z} \;  \gamma_{-k\,\alpha})$, $H_{\text M}$  looks like
\begin{align}
 H_{\text M}&=E_0-\frac{1}{2}\sum_k\left(\Lambda_{k\,z}+2\,\Lambda_{k\,\alpha}\right)
+ \frac{1}{2}\sum_k \Phi_k^\dagger\, H_k^d \,\Phi_k\,,\nonumber \\
H_k^d&=\smat{\Omega_k&0\\0&\Omega_k},\qquad  \Omega_k=\smat{\omega_{k\,z}&0\\0&\omega_{k\,\alpha} },\qquad
\Phi_k=T_k\,\Psi_k\,, \nonumber \\
  T_k&=\smat{u_k&v_k\\v_k&u_k}, \qquad
u_k=\frac{1}{\sqrt 2}\smat{\sqrt{1+\frac{\Lambda_{k\,z}}{\omega_{k\,z}}}&0\\
0&\sqrt{1+\frac{\Lambda_{k\,\alpha}}{\omega_{k\,\alpha}}}}, \qquad
v_k= \frac{1}{\sqrt 2}\smat{\sqrt{-1+\frac{\Lambda_{k\,z}}{\omega_{k\,z}}}&0\\
0&\sqrt{-1+\frac{\Lambda_{k\,\alpha}}{\omega_{k\,\alpha}}}}. \nonumber
\end{align}
$H_{\text M}$ can further be expressed as
\be
H_{\text M}=E_0-\frac{1}{2}\sum_k\left(\Lambda_{k\,z}
+2\Lambda_{k\,\alpha}-\omega_{k\,z}-2\omega_{k\,\alpha}\right)
+\sum_k\left(\omega_{k\,z}\;\gamma_{k\,z}^\dagger \gamma_{k\,z}
+ \omega_{k\,\alpha}\;\gamma_{k\,\alpha}^\dagger \gamma_{k\,\alpha}\right).
\ee
Therefore, it turns out that $\omega_{k\,z}$ and $\omega_{k\,\alpha}$ are like the non-degenerate
longitudinal and doubly-degenerate transverse branches of triplet dispersion relations, respectively.
When $\Delta=1$, the two branches merge to each other leading to a triply-degenerate
single triplet branch.
The parameters $\mu$, $\bar s$, $P_z$, $P_\alpha$, $Q_z$ and $Q_\alpha$ are determined
by solving the six saddle-point equations:
\bea
\left\langle \frac{\pd H_{\text M}}{\pd \mu}\right\rangle=0,\quad
\left\langle \frac{\pd H_{\text M}}{\pd \bar s}\right\rangle=0,\quad
\left\langle \frac{\pd H_{\text M}}{\pd  P_z}\right\rangle=0,\quad
\left\langle \frac{\pd H_{\text M}}{\pd  P_\alpha}\right\rangle=0,\quad
\left\langle \frac{\pd H_{\text M}}{\pd  Q_z}\right\rangle=0,\quad
\left\langle \frac{\pd H_{\text M}}{\pd  Q_\alpha}\right\rangle=0,\nonumber
\eea
which lead to the following six self-consistent equations at $T=0$~K.
\begin{align}
\mu&=\frac{J_2}{2N}\sum_k\left(\Delta\,\frac{2\Delta_{k\,z}-\Lambda_{k\,z}}{\omega_{k\,z}}
+2\;\frac{2\Delta_{k\,\alpha}-\Lambda_{k\,\alpha}}{\omega_{k\,\alpha}} +\Delta+2 \right)\cos(ka)
-J_1\left(\frac{\Delta}{4}+\frac{1}{2} \right),\nonumber \\
\bar s ^2&=\frac{5}{2}-\frac{1}{N}\sum_k\left( \frac{\Lambda_{k\,z}}{\omega_{k\,z}}+
2\;\frac{\Lambda_{k\,\alpha}}{\omega_{k\,\alpha}}  \right),\nonumber \\
P_z&=\frac{1}{N}\sum_k\left(  \frac{\Lambda_{k\,z}}{\omega_{k\,z}}-1\right)\cos(ka), \nonumber\\
P_\alpha&=\frac{1}{N}\sum_k\left(  \frac{\Lambda_{k\,\alpha}}{\omega_{k\,\alpha}}-1\right)\cos(ka),\nonumber \\
Q_z&=-\frac{2}{N}\sum_k   \frac{\Delta_{k\,z}}{\omega_{k\,z}}\cos(ka),\nonumber \\
Q_\alpha&=-\frac{2}{N}\sum_k   \frac{\Delta_{k\,\alpha}}{\omega_{k\,\alpha}}\cos(ka).
\label{selfconsistent}
\end{align}
For fixed values of $J_1$, $J_2$ and $\Delta$, the six self-consistent solutions are obtained
from equations~(\ref{selfconsistent}) and are employed to determine the dispersion relations,
ground state energy, spin gap and dimer order.
For $J_2=0$, values of the parameters, $P_z$, $P_\alpha$, $Q_z$ and $Q_\alpha $ must be zero,
and they are non-zero when  $J_2\neq 0$.
The solutions for $\mu$ are always negative while those for  $\bar s^2$ are always positive.
These six self-consistent equations are found to converge
in most of the anisotropic parameter regions except the regime where spin gap is
vanishingly small which occurs when $J_1\approx J_2$ and $\Delta\approx 0$.
So, the values of ground state energy, spin gap and dimer order
in this regime are not plotted in the respective figures in subsequent sections.
The ground state energy per site ($E_{\rm G}$) is given by the following expression,
\[
E_{\rm G}=\frac{E_0}{N}-\frac{1}{2N}\sum_k\left(\Lambda_{k\,z}
+2\Lambda_{k\,\alpha}-\omega_{k\,z}-2\omega_{k\,\alpha}\right).
\]
For $J_1= J_2$ and $\Delta=1$, $E_{\rm G}=-0.45130123J_1$, which is only 0.18\% lower than the
exact Bethe-Ansatz result,  i.e., $(0.25-\ln 2)J_1=-0.44314718J_1$.
The values of $E_{\rm G}$ are very close to the exact diagonalization results in the entire parameter regime
except the point $\Delta=0$ and those are shown in figure~\ref{genergy}.

The expression of ground state dimer order looks like
\begin{align}
\langle \mathcal{O}_{\rm D} \rangle &=
D_0-\frac{1}{2}\sum_k\left(X_{k \,z}+2X_{k\,\alpha }
+ \frac{\Lambda_{k\,z} X_{k \,z}-4\, \Delta_{k\,z} Y_{z}}{\omega_{k\,z}}
+2\,\frac{\Lambda_{k\,\alpha} X_{k \,\alpha}-4\, \Delta_{k\,\alpha} Y_{\alpha}}{\omega_{k\,\alpha}}\right), \nonumber \\
D_0&=-\frac{3}{4}\bar s^2+P_zP_\alpha -Q_zQ_\alpha+\frac{1}{2}\left(P_\alpha ^2-Q_\alpha^2 \right), \nonumber\\
X_{k \,z}&=\frac{1}{4}\left[ 1+2(\bar s^2-2P_\alpha)\cos(ka) \right],\qquad
X_{k \,\alpha}=\frac{1}{4}\left[ 1+2(\bar s^2-P_z-P_\alpha)\cos(ka) \right],\nonumber \\
Y_z&=\frac{1}{4}\left( \bar s^2+2Q_\alpha \right),\quad
Y_\alpha=\frac{1}{4}\left( \bar s^2+Q_z+Q_\alpha \right).\nonumber
\end{align}
The values of this $\mathcal{O}_{\rm D}$ along with plaquette and exact diagonalization results
 are shown in figure~\ref{do}. Bond operator results are discontinued at those
points where convergence fails to be attained.

\begin{figure}[!b]
\begin{center}
\psfrag{J2}{$J_2/J_1$}
\psfrag{p1}{\text{\scriptsize{$-\piup$}}}
\psfrag{p2}{\text{\scriptsize{$-\frac{\piup}{2}$}}}
\psfrag{p3}{\text{\scriptsize{$0$}}}
\psfrag{p4}{\text{\scriptsize{$\frac{\piup}{2}$}}}
\psfrag{p5}{\text{\scriptsize{$\piup$}}}
\psfrag{1.3}{\text{\tiny{$1.3$}}}
\psfrag{0.9}{\text{\tiny{$0.9$}}}
\psfrag{0.5}{\text{\tiny{$0.5$}}}
\psfrag{0.1}{\text{\tiny{$0.1$}}}
\psfrag{1.1}{\text{\tiny{}}}
\psfrag{0.7}{\text{\tiny{}}}
\psfrag{0.3}{\text{\tiny{}}}
\psfrag{1}{\text{\scriptsize{$J_2=-0.5$}}}
\psfrag{11}{\text{\scriptsize{$\Delta=0.00$}}}
\psfrag{21}{\text{\scriptsize{$\Delta=0.25$}}}
\psfrag{31}{\text{\scriptsize{$\Delta=0.50$}}}
\psfrag{41}{\text{\scriptsize{$\Delta=0.75$}}}
\psfrag{5}{\text{\scriptsize{$J_2=0.5,\,\Delta=0.00$}}}
\psfrag{6}{\text{\scriptsize{$J_2=0.5,\,\Delta=0.25$}}}
\psfrag{7}{\text{\scriptsize{$J_2=0.5,\,\Delta=0.50$}}}
\psfrag{8}{\text{\scriptsize{$J_2=0.5,\,\Delta=0.75$}}}
\psfrag{w}{$\omega_k$}
\psfrag{k}{\text{\scriptsize{$k$}}}
\psfrag{alpha}{\text{\scriptsize{$\omega_{k\,\alpha}$}}}
\psfrag{zzzzz}{\text{\scriptsize{$\omega_{k\,z}$}}}
\includegraphics[scale=0.29]{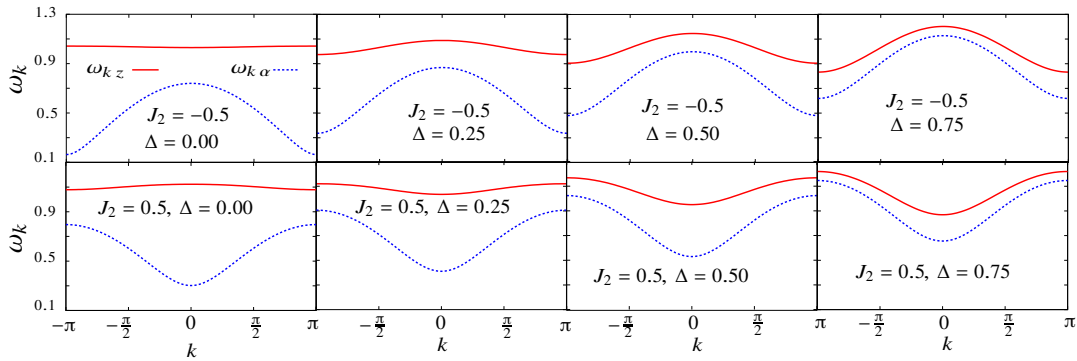}
\caption{\label{dispersion}{(Color online) Dispersion relations (in unit of $J_1$)
with respect to $k$. $\omega_{k\,z}$  and $\omega_{k\,\alpha}$
are in  red (solid) and blue (dashed) lines, respectively.}}
\end{center}
\end{figure}
\begin{figure}[!b]
\vspace{-3mm}
\begin{center}
\psfrag{Delta}{\text{\scriptsize{$\Delta$}}}
\psfrag{do}{\text{\scriptsize{$\mathcal O_{\rm D}$}}}
\psfrag{sz}{\text{\scriptsize{$\mathcal O^z_{\rm S}$}}}
\psfrag{sx}{\text{\scriptsize{$\mathcal O^x_{\rm S}$}}}
\psfrag{gap}{$E_{\rm Gap}/J_1$}
\psfrag{J2byJ1}{$J_2/J_1$}
\psfrag{a}{$\Delta=0.0$}
\psfrag{Diagonalization}{\scriptsize Diagonalization}
\psfrag{Bond Operator}{\scriptsize Bond Operator}
\psfrag{b}{$\Delta=0.3$}
\psfrag{c}{$\Delta=0.7$}
\psfrag{d}{$\Delta=1.0$}
\includegraphics[scale=1.05]{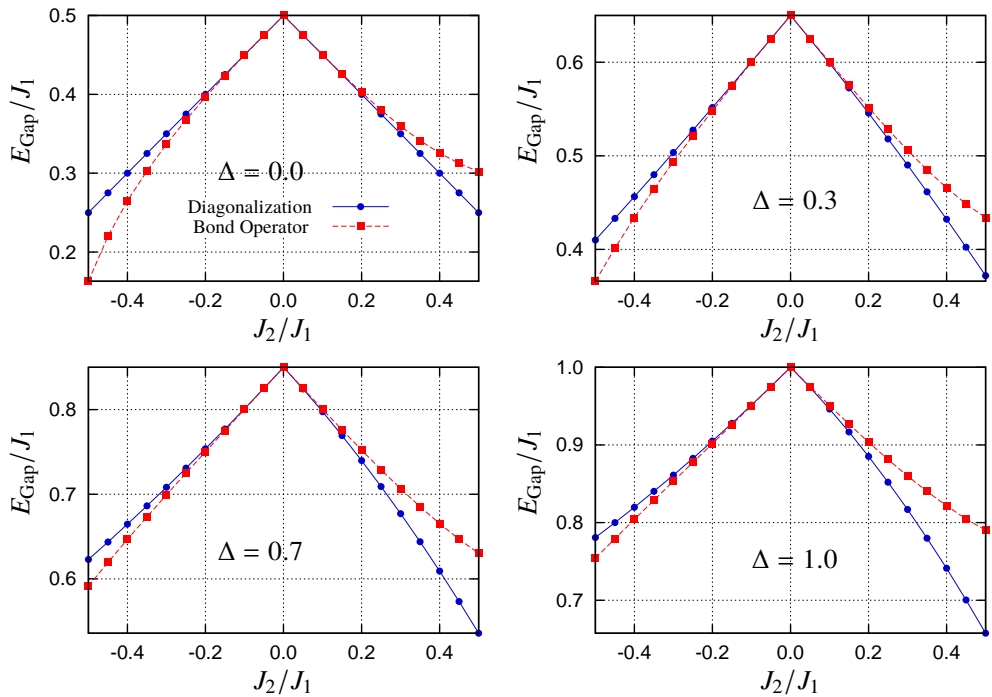}
\caption{(Color online)
Plots of $E_{\rm Gap}/J_1$ with respect to $J_2/J_1$.
Blue (solid line, circle) and red (dashed line, square) correspond
to the exact diagonalization and bond operator results, respectively. }
\label{spingap}
\end{center}
\end{figure}

The dispersion relations, ${\omega_{k\,z}}$ and
$\omega_{k\,\alpha}$ arising from the excitations of spin-triplet states
are shown in figure~\ref{dispersion}.
$J_2=0$ is a dispersionless point, where both  ${\omega_{k\,z}}$ and $\omega_{k\,\alpha}$
are flat since energy propagation is impossible in
the absence of inter-bond interaction $J_2$. This particular point
is not shown in figure~\ref{dispersion}.
For $\Delta=0$, ${\omega_{k\,z}}$ is almost flat with small curvature,
concave down for AFM $J_2$ while concave up for FM $J_2$,
 whereas, $\omega_{k\,\alpha}$ has the maximum curvature.
 ${\omega_{k\,z}}$ will be perfect dispersionless if the part of
the Hamiltonian [equation~(\ref{bondham})] containing four-$t$ operators is neglected.
It is interesting to note that even though $\Delta=0$,
${\omega_{k\,z}}$ is non-zero. So, it establishes the fact that the
existence of longitudinal mode is quantum mechanically possible in the absence of the
longitudinal part of the Hamiltonian. These modes hardly participate
in energy propagation.
However, in general, both ${\omega_{k\,z}}$ and
$\omega_{k\,\alpha}$ are concave up for AFM $J_2$
and concave down for FM $J_2$. Bandwidth for  ${\omega_{k\,z}}$
($\omega_{k\,\alpha}$) increases (decreases) with increasing $\Delta$ for fixed value of $J_2$.
On the other hand, bandwidths for  ${\omega_{k\,z}}$
 as well as $\omega_{k\,\alpha}$ increase with increasing value of both
AFM and FM  $J_2$ for fixed value of $\Delta$.
However, for a fixed value of AFM $J_2$, bandwidths for ${\omega_{k\,z}}$ and
$\omega_{k\,\alpha}$ are separately the same to them for that value of FM $J_2$ for a definite value of $\Delta$.
The minima of triplet dispersion relations, ${\omega_{k\,z}}$ and
$\omega_{k\,\alpha}$ are found at $k=0$ when $J_2>0$ and at  $k=\piup$ when $J_2<0$ as far as
$|J_2|\leqslant J_1$. So, the value of spin gap can be estimated by using the relations,
$E_{\rm Gap}=\min\left[\omega_{k\,z}(k=0),\omega_{k\,\alpha}(k=0)\right]$ when $J_2>0$ and
$E_{\rm Gap}=\min\left[\omega_{k\,z}(k=\piup),\omega_{k\,\alpha}(k=\piup)\right]$ when $J_2<0$.
Since $\omega_{k\,\alpha}\leqslant \omega_{k\,z}$ for any value of the wave vector, $k$, in the
anisotropic parameter regime $0\leqslant \Delta\leqslant 1$,
$E_{\rm Gap}=\omega_{k\,\alpha}(k=0)$ for AFM $J_2$ and
$E_{\rm Gap}=\omega_{k\,\alpha}(k=\piup)$ for FM $J_2$.
Figure~\ref{dispersion} shows that the value of $E_{\rm Gap}$ increases with the increase of
$\Delta$ in every case. Variation of $E_{\rm Gap}$ with $J_2/J_1$ is shown in
figure~\ref{spingap} along with the exact diagonalization results.
$E_{\rm Gap}$ is found to decrease in the absence of the part of equation~(\ref{bondham}) containing four-$t$
operators terms.

\section{Jordan-Wigner representation}
\label{J-W}
This model is exactly solvable in terms of Fermi gas of
spinless fermions for $\Delta=0$ by
using the Jordan-Wigner transformation \cite{J-W}
\begin{align}
S_i^+&=c_i^\dagger \re^{\ri\piup\sum_{j=1}^{i-1}\hat n_j},\nonumber \\
S_i^-&= \re^{-\ri\piup\sum_{j=1}^{i-1}\hat n_j}c_i\,,\nonumber \\
S_i^z&=\hat n_i -\frac{1}{2}\,,\nonumber
\end{align}
where $c_i$ and $c_i^\dagger$ are the spinless fermion annihilation and
creation operators, respectively.
$\hat n_i = c_i^\dagger c_i$ is the usual fermion number operator.
This bond alternating system has a translational symmetry of two lattice
units and so it becomes useful to introduce
two types of spinless fermions defined on
odd and even lattice sites by relabeling them as:
$c_i=a_i$ for odd sites and $c_i=b_i$ for even sites. As a result,
the Hamiltonian becomes
\begin{align}
H&=\sum_{i=1,3,5,\ldots}\left[\frac{J_1}{2}\left(a_i^\dagger b_{i+1}+ b^\dagger_{i+1}a_i\right)
+J_1\Delta\left(a_i^\dagger a_i-\frac{1}{2}\right)
\left(b_{i+1}^\dagger b_{i+1}-\frac{1}{2}\right)\right]\nonumber\\
&+
\sum_{i=1,3,5,\ldots}\left[\frac{J_2}{2}\left(b_{i+1}^\dagger a_{i+2}+a_{i+2}^\dagger b_{i+2}\right)
+J_2\Delta\left(b_{i+1}^\dagger b_{i+1}-\frac{1}{2}\right)
\left(a_{i+2}^\dagger a_{i+2}-\frac{1}{2}\right)\right].
\label{jwham}
\end{align}
For $\Delta\neq 0$, four-operator terms
can be treated by the mean-field analysis.
By allowing contractions of types, say, $C_a=\langle a_i^\dagger b_{i+1}\rangle$ and
$C_b=\langle b_i^\dagger a_{i+1}\rangle$,
 the mean-field
Hamiltonian in momentum space reads as
\be
H_{\rm MF}=\sum_k
\left[C_k \,a_k^\dagger b_k+\bar C_k\, b^\dagger_k a_k
+h\left( a_k^\dagger a_k-b^\dagger_k b_k\right) \right]
+\frac{N\Delta}{8}(J_1+J_2)+\frac{N\Delta}{2}\left(J_1|C_a|^2+J_2|C_b|^2 \right),
\ee
where $C_k=(\frac{J_1}{2}-J_1\,\Delta\, C_a)\,\re^{-\ri ka}+(\frac{J_2}{2}-J_2\,\Delta\, C_b)\, \re^{\ri ka}$ and
$h=\frac{\Delta}{2}(J_1-J_2)$.
In this fermionic description, $h$ acts as a chemical potential whose value is the same
for every particle but opposite in sign for two different kinds of particles, say,
positive for $a$ and negative for $b$.
In other words, $h$ acts as a staggered field giving rise to
a periodic potential experienced by the particles with the periodicity of two
lattice units. As a result,
Brillouin zone shrinks to its half yielding a spin gap
in its boundary. On the other hand, $h$ vanishes for the uniform bond strength,
i.e., when $J_1 =J_2$ and so the spin gap.

By using the fermionic Bogoliubov transformation
\be
a_k=u_k\,\alpha_k+v_k\,\beta_k\,,\qquad
b_k=-v_k^* \,\alpha_k+u_k^* \,\beta_k\,,\nonumber
\ee
where $u_k=r\,\re^{\ri\theta_k}$, $v_k=r'\re^{\ri\theta_k}$ and
$\re^{2\ri\theta_k}=C_k/|C_k|$,
the diagonalized mean-field Hamiltonian looks like
\be
H_{\rm MF}=\sum_k E(k)
\left(\alpha_k^\dagger \alpha_k-\beta_k^\dagger \beta_k\right)+\frac{N\,\Delta}{8}\,
\left(J_1+J_2\right)+\frac{N\Delta}{2}\left(J_1|C_a|^2+J_2|C_b|^2 \right),
\ee
where $E(k)=\sqrt{h^2+|C_k|^2}$.

By allowing contractions of other possible ways, say, $D_a=\langle a_i^\dagger a_i-\frac{1}{2}\rangle$ and
$D_b=\langle b_{i+1}^\dagger b_{i+1}-\frac{1}{2}\rangle$,
 the mean-field
Hamiltonian in momentum space becomes
\begin{align}
H_{\rm MF}&=\sum_k
\left[D_k \,a_k^\dagger b_k+\bar D_k\, b^\dagger_k a_k
+\Delta\,(J_1+J_2)\left(D_b\, a_k^\dagger a_k+D_a\,b^\dagger_k b_k\right) \right]\nonumber \\
&\quad-\frac{N}{4}\Delta(J_1+J_2)\left(D_a+D_b+2D_aD_b \right),
\end{align}
where $D_k=\frac{1}{2}\left(J_1 \,\re^{-\ri ka}+J_2 \,\re^{\ri ka}\right)$.
By performing the same Bogoliubov transformation, diagonalized
Hamiltonian reads as
\be
H_{\rm MF}=\sum_k \omega(k)
\left(\alpha_k^\dagger \alpha_k-\beta_k^\dagger \beta_k\right)-\frac{N}{2}\,D\,h, 
\ee
where $\omega(k)=\sqrt{h^2+|D_k|^2}$ and $h=-\Delta (J_1+J_2)D$, when
$D_a=-D_b=D$.

The mean-field parameters, $C_a$, $C_b$, and $D$ will be determined by solving
self-consistent equations defined in the four different phases \cite{Verkholyak}.

i) Paramagnetic (homogeneous) phase: when $C_a=C_b$ \cite{Bulaevskii},
\be
C_a=-\frac{1}{N}\sum_{k} \frac{\cos^2(ka)}{E(k)}\;\left[
\left( \frac{J_1+J_2}{2}\right)(1-2\,\Delta\, C_a)\right]
\left[n_\beta(k)-n_\alpha(k)\right].
\ee

ii) Staggered AFM (SAFM) phase: when $C_a=-C_b$,
\be
C_a=\frac{1}{N}\sum_{k} \frac{\sin^2(ka)}{E(k)}\;\left[
\left( \frac{J_1-J_2}{2}\right)(1-2\,\Delta\, C_a)\right]
\left[n_\beta(k)-n_\alpha(k)\right].
\ee

iii) Alternating NN hopping (dimer) phase: when
$C_a=\eta+\delta$ and $C_b=\eta-\delta$,
\bea
\eta&=&-\frac{1}{N}\sum_{k} \frac{\cos^2(ka)}{E(k)}\;\left[
\left( \frac{J_1+J_2}{2}\right)(1-2\Delta \eta)-\delta \Delta (J_1-J_2)\right]
\left[n_\beta(k)-n_\alpha(k)\right],\nonumber\\
\delta&=&\frac{1}{N}\sum_{k} \frac{\sin^2(ka)}{E(k)}\;\left[
\left( \frac{J_1-J_2}{2}\right)(1-2\Delta \eta)-\delta \Delta (J_1+J_2)\right]
\left[n_\beta(k)-n_\alpha(k)\right].
\label{self-consistent}
\eea

iv) Uniform AFM (UAFM) phase: when $D_a=-D_b$,
\be
1=\frac{1}{N}\sum_k\frac{\Delta(J_1+J_2)}{\omega(k)}\left[n_\beta(k)-n_\alpha(k)\right].
\ee
 $D_a$ vanishes when $J_1=-J_2$.
Another choice $D_a=D_b$, which corresponds to the uniform FM phase gives rise to $D_a=0$.
So, this choice does not produce a non-trivial result, and thus deserves no further attention.
$n_\alpha(k)=\langle \alpha_k^\dagger \alpha_k \rangle$
and $n_\beta(k)=\langle \beta_k^\dagger \beta_k\rangle$
are the fermionic occupation probabilities at
temperature~$T$. At zero temperature, only the negative energy states
are filled up, so, $n_\alpha(k)=0$ and $n_\beta(k)=1$.
In this situation, the expressions of $E_{\rm G}$ can be written down
easily for each mean-field case. For example, in the dimer phase (iii), it is given by
\[
E_{\rm G}=-\frac{1}{\piup} \int_0^{\piup/2} E_k\,\rd k+\frac{\Delta}{2}
\left[J_1(\eta+\delta)^2+J_2 (\eta-\delta)^2\right]
+\frac{\Delta}{8}\,\left(J_1+J_2\right).
\]
For  $\Delta=0$, ground state energy
can be exactly evaluated for any values of both $\frac{J_2}{J_1}$
and temperatures. For example, at $T=0$,
$E_{\rm G}=-\frac{1}{\piup}J_1=-0.31830989J_1$, when
$\frac{J_2}{J_1}=1$ and $\Delta=0.$

This mean-field ground state energy has been improved
 by considering the second order contribution attributed to
the fluctuations around the mean field. This correction may be evaluated
by using the standard expression at $T=0$ \cite{Wang},
\be
\Delta E= \sum_f\frac{\left|\langle g\left|\left(H-H_{\rm MF} \right)
\right|f\rangle \right|^2}{E_g-E_f}\,,
\label{Correction}
\ee
where $|g\rangle=\prod_k \beta_{k}^\dagger |0\rangle$ and
$|f\rangle=\alpha_{k_1+q}^\dagger \alpha_{k_2-q}^\dagger |0\rangle$ are the
ground state and the excited states of $H_{\rm MF}$, respectively.
The state $|f\rangle$ has two excited particles at wave vectors
$k_1+q$ and $k_2-q$ at the positive energy branch. Non-zero
contributions come from the four-operator terms in equation~(\ref{jwham}).
Numerical evaluation of equation~(\ref{Correction})
leads to $\Delta E=-0.0171 J_1$ per site  for $\frac{J_2}{J_1}=1$
and $\Delta=1$. The final value
of the ground state energy is $E_{\rm G}(\Delta=1,J_1=J_2)=-0.4367J_1$, after
the second order correction which is very close to the
exact Bethe-Ansatz result, i.e., $-0.4431J_1$.

The mean-field ground state energies have been plotted along with the
exact diagonalization and bond operator results in the figure~\ref{genergy}.
Different Jordan-Wigner based mean-field results are
UAFM (dark-cyan, dashed-dot line, pentagon),
homogeneous (black, dashed-dot line, diamond),
SAFM (green, dashed line, triangle) and
dimer (purple, dotted line, inverted triangle).
For $\Delta=0$, Jordan-Wigner representation provides the
exact ground state energy and thus it coincides
with the exact diagonalization result (figure~\ref{genergy}).
However, when $\Delta>0$, SAFM and UAFM phases do not at all agree with
the exact diagonalization results.
$E_{\rm G}$ evaluated in the UAFM phase is always higher, while
that evaluated in SAFM phase is higher (lower) when $J_2$ is AFM (FM).
Dimer and homogeneous phases do agree with
the exact diagonalization result only in the AFM  $J_2$ region.
On the other hand, $E_{\rm G}$ evaluated in the bond operator formalism
mostly coincides with the exact diagonalization result
apart from the point $\Delta=0$.
For $\Delta=0$, $E_{\rm G}$ derived in this formalism
coincides with the exact diagonalization result around $|J_2/J_1|\approx 0$.

\section{Exact diagonalization results}
\label{E-D}
\begin{figure}[!t]
\begin{center}
\psfrag{J2}{$J_2/J_1$}
\psfrag{a}{$\Delta=0.0$}
\psfrag{b}{$\Delta=0.3$}
\psfrag{c}{$\Delta=0.7$}
\psfrag{d}{$\Delta=1.0$}
\psfrag{energy}{Energy/$J_1$}
\includegraphics[scale=0.95]{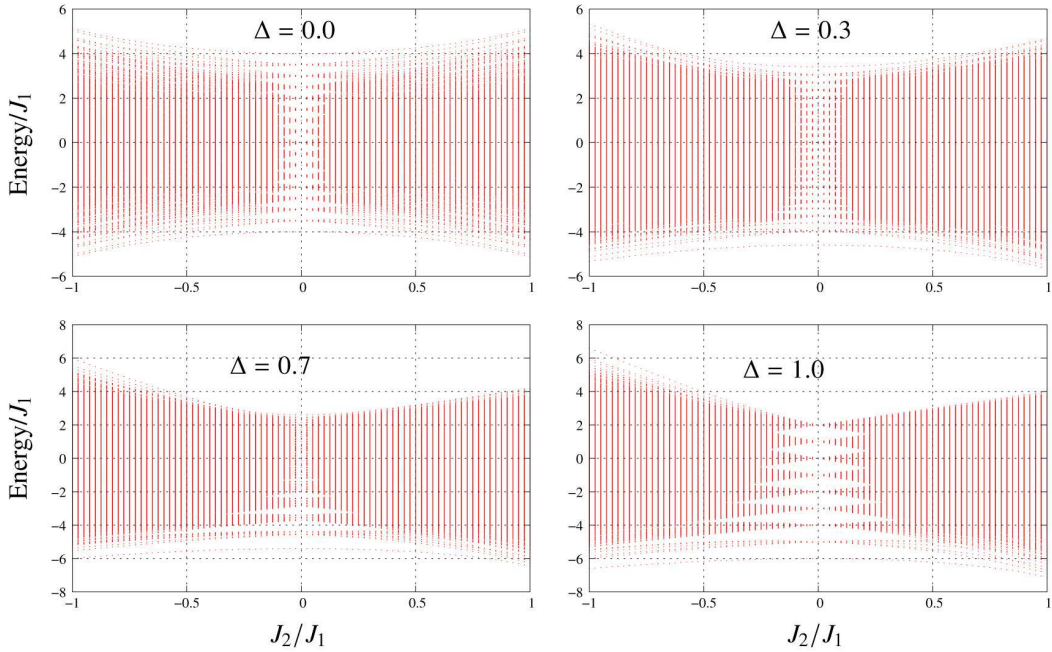}
\caption{\label{energy-spectra}{(Color online) Plot of all energies (in unit of $J_1$)
with respect to $J_2/J_1$ for $\Delta=0.0$, $\Delta=0.3$,
$\Delta=0.7$, $\Delta=1.0$ and $N=16$. With the
increasing $\Delta$, the width of the energy band
increases and moves toward the low energy side.}}
\end{center}
\end{figure}

The ground state energy, spin gap and several ground state correlation functions
have been obtained numerically at zero temperature. Ground state energy has been
compared with the theoretical results.
The spin gap is defined as the difference between the energies of
ground state and the lowest excited state
for a chain of finite number of spins.
The Lanczos exact diagonalization technique is the most suitable
algorithm  when a few extreme eigenvalues are required.
To find the ground state energy,
the Hamiltonian is diagonalized in a subspace where $S_{\rm T}^z=0$.
 The Hilbert space is further reduced
by exploiting two different symmetries of this Hamiltonian.
The first one is the translational invariance of two lattice
units while the second one is the spin inversion in every site.
Due to the spin inversion symmetry, the energy eigenvalues satisfy the
relation, $E(S_{\rm T}^z)=E(-S_{\rm T}^z)$.
The periodic boundary condition is taken into account in every case.
As a result, two different momentum wave vectors, $qT2$ and $qR$ are
defined to associate the symmetries of Hamiltonian with
the translation of two lattice units and the spin inversion, respectively.
Eventually, including those symmetries in
the modified Lanczos algorithm \cite{Lanczos},
this computational procedure could find the eigenenergies of the
spin chain up to the length ($N$) of 32 sites. The ground state is unique and
corresponds to the wave vectors $qT2=0$ and $qR=\piup\; modulo\;(N,4)$ for
both AFM and FM $J_2$.
The doubly degenerate lowest excited state
corresponds to the $S_{\rm T}^z=\pm 1$ and $qT2=0$ but $qR=0$ for
AFM $J_2$ while  $qR=\frac{4\piup}{N}\; quotient\; (N,4)$ for FM $J_2$.
The Hamiltonian, $H$, [equation~(\ref{ham})] exhibits another useful symmetry
in which the unitary operator,
$U=\prod_j \exp(\ri\piup jS_j^z)$ transforms
$H$ as $UH(J_1,J_2,\Delta)U^\dagger=H(-J_1,-J_2,-\Delta)$.
This symmetry transformation leads to the following result:
 when $\Delta=0$, $UH(J_1,J_2,\Delta=0)U^\dagger
=-H(J_1,J_2,\Delta=0)$. So, energy spectrum of $H$ has the
reflection symmetry around the zero energy.
This symmetry is observed in the energy spectrum for $\Delta=0$
and is shown in figure~\ref{energy-spectra}.
The spectrum for $\Delta=0$ is also symmetric around the point $J_2=0$,
although no transformation is found to justify this symmetry.
Obviously those symmetries are lost when $\Delta \neq 0$.

The full energy spectra of this model for four different values of
$\Delta$ are plotted with respect to $J_2/J_1$ and are shown in
figure~\ref{energy-spectra}
in which the uniqueness of the ground state and finite spin gap has been
observed clearly. For $\Delta=0$, the spectrum is symmetric around zero energy,
but the spectra move toward low energy side and
at the same time symmetry is lost when $\Delta \neq 0$.
The spectra are found to split into several bands around $J_2=0$.
The number of bands increases with a
decreasing $\Delta$. The nature of those energy spectra remains unaltered in
the open boundary condition.

To examine the effect of non-uniformity of the alternating bond strength
 on the spin gap,
the modified Lanczos algorithm is employed
designed for finite-size spin chain having integral multiple of 4,
$N=16, 20,  \ldots,  32$.
Ground state energy per site as well as
spin gap depend on both the
chain length ($N$) and the relative
difference between alternating bond strengths, i.e.,
$\delta=(J_1-J_2)/J_1$.
The spin gap is defined as
\be
E_{\rm Gap}(N,\Delta,\delta)=E_{\rm F}\left(N,\Delta,\delta,S_{\rm T}^z=\pm 1\right)
-E_{\rm Gr}\left(N,\Delta,\delta,S_{\rm T}^z=0\right),
\label{energygap}
\ee
where $E_{\rm Gr}$ and $E_{\rm F}$ are the ground state and the first excited state
energies, respectively.
For $\delta=0$, the spectrum is gapless for the entire range of
$0\leqslant \Delta \leqslant 1$, for AFM $J_2$ whereas,
spin gap is found to develop as soon as $\delta \neq 0$
for the same range of $\Delta$ and AFM $J_2$. Thus, $\delta=0$ serves as the
critical point for this transition. On the other hand, for FM  $J_2$, this spin gap
is observed for any value of $\delta$ and $\Delta$.
The spin gap has been estimated by
using several values of $\delta$ and $\Delta$
within the range $0 < \delta \leqslant0.10$ and $0 < \Delta \leqslant 1.0$
for the chain lengths those are integral multiple of 4, i.e.,
$N=16,20, \ldots, 32$.

\begin{figure}[!t]
\begin{center}
\psfrag{Delta}{\text{\scriptsize{$\Delta$}}}
\psfrag{do}{\text{\scriptsize{$\mathcal O_{\rm D}$}}}
\psfrag{sz}{\text{\scriptsize{$\mathcal O^z_{\rm S}$}}}
\psfrag{sx}{\text{\scriptsize{$\mathcal O^x_{\rm S}$}}}
\psfrag{spingap}{\text{\scriptsize{$E_{\rm Gap}/J_1$}}}
\psfrag{J2}{\text{\scriptsize{$J_2/J_1$}}}
\psfrag{a}{\scriptsize (a)}
\psfrag{b}{\scriptsize (b)}
\psfrag{c}{\scriptsize(c)}
\psfrag{d}{\scriptsize(d)}
\includegraphics[scale=0.6]{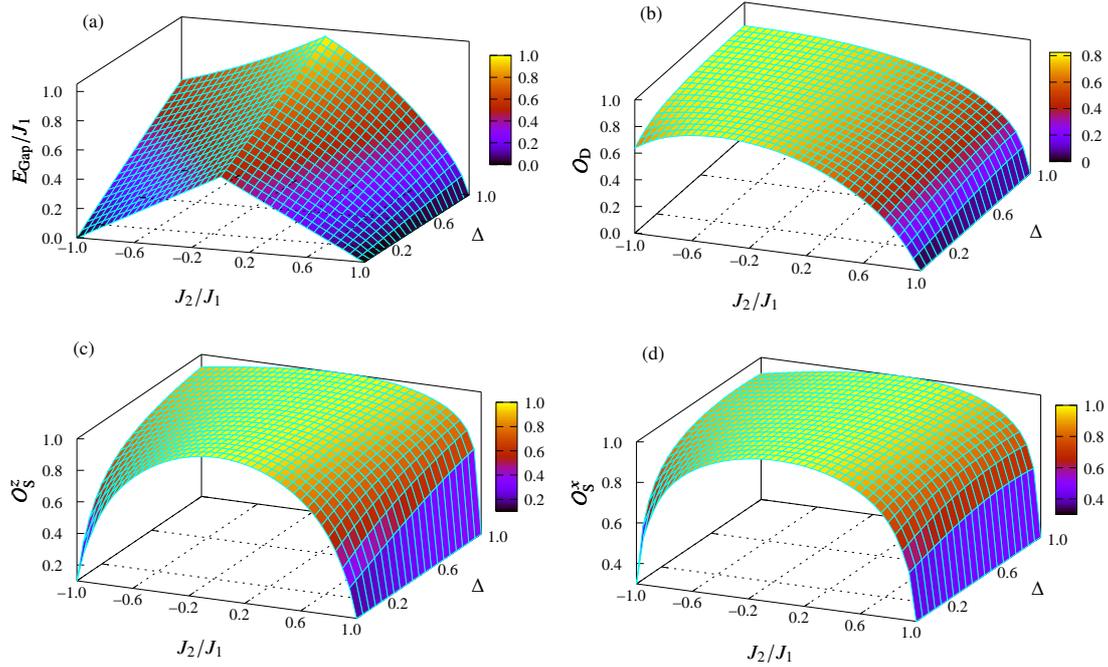}
\caption{\label{ground_order}{(Color online)
Three dimensional plots of $E_{\rm Gap}/J_1$ (a), $\mathcal O_{\rm D}$ (b),
$\mathcal O^z_{\rm S}$ (c) and $\mathcal O^x_{\rm S}$ (d)
 with respect to both $J_2/J_1$ and $\Delta$.}}
\end{center}
\end{figure}

To obtain the values of $E_{\rm G}$ and $E_{\rm Gap}$ in large $N$
limit, finite size extrapolations have been performed by
using the Vanden-Broeck-Schwartz (VBS) algorithm \cite{VBS}
with $\alpha_{\rm VBS}=-1$ in addition to the
Bulirsch-Stoer (BST) algorithm \cite{BST}.
Comparisons of those estimates with theoretical results reveal that
the VBS algorithm yields more accurate values for both
$E_{\rm G}$ and $E_{\rm Gap}$ than the BST algorithm.
For the extrapolations, the values of $E_{\rm G}$ and $E_{\rm Gap}$ for chains
of five different lengths like $N=16,20, \ldots, 32$
are considered. The extrapolated value of   $E_{\rm G}$
agrees with the exact result up to the sixth decimal positions.
For example, when $\Delta=1$ and $\delta=0$, the extrapolated value of
ground state energy per site
is  $-0.44314728 J_1$ which is extremely close to the exact Bethe-Ansatz value,
$-0.44314718J_1$ or thus only $0.0000225\%$
lower than the exact value. On the other extreme point, i.e.,
when $\Delta=0$ and $\delta=0$,
the extrapolated value of $E_{\rm G}$
is  $-0.31830988 J_1$ which completely agrees with the exact value
 $-J_1/\piup=-0.31830988J_1$. Therefore, it is expected that the accuracy of
those numerical estimations is very high.
The extrapolated values of $E_{\rm Gap}$ are found by using the VBS algorithm
and are plotted in figure~ \ref{ground_order}~(a).
This three-dimensional plot reveals that
$E_{\rm Gap}$ vanishes over the line $J_2/J_1=1$  and
on the point, $J_2/J_1=-1$, $\Delta=0$.
$E_{\rm Gap}$ is found to increase with the increase of both $\delta$
and $\Delta$ up to the line $J_2/J_1=0$. However, it again decreases toward
FM region. The magnitude of spin gap is symmetric around $J_2=0$ for
$\Delta=0$, due to the symmetry of energy spectrum.

\section{Ground state properties}
\label{ground-properties}
The results obtained using various methods in the
previous sections are summarized here.
A comparison of values of $E_{\rm G}$ obtained in exact diagonalization, bond operator
formalism and various Jordan-Wigner based mean-field methods has been displayed in
figure~\ref{genergy}. It shows that the values of $E_{\rm G}$ obtained in
exact diagonalization and bond operator based mean-field
formalism agree remarkably when $\Delta>0$ while
Jordan-Wigner based mean-field methods totally disagree.
On the other hand, for $\Delta=0$, Jordan-Wigner based exact result
coincides with the exact diagonalization value but the results obtained by bond operator
formalism show a qualitative agreement. Thus, it reveals that
these two different analytic formalisms, bond operator and Jordan-Wigner,
predict the true values of $E_{\rm G}$ in two different regions for these
bond alternating models.

Ground state expectation value of the dimer order, $\mathcal{O}_{\rm D}$, has been
evaluated numerically.
 In this expression, the stronger AFM bonds ($J_1$) are assumed between
the sites $i$ and $i+1$ while the FM or weaker AFM bonds ($J_2$) are
acting between the sites $i+1$ and $i+2$. The variation of
$\mathcal{O}_{\rm D}$ with respect to both $\Delta$ and $J_2/J_1$
has been shown in figure~\ref{ground_order}~(b).
The values of $\mathcal{O}_{\rm D}$ obtained by using
exact diagonalization, bond operator formalism and four-spin plaquette
have been shown in figure~\ref{do}. All the methods show a good qualitative agreement.
However, the bond operator based results do not vanish over the line $J_1=J_2$.
The exact diagonalization results
quite agree with the DMRG results reported earlier by
Watanabe and Yokoyama for $\Delta=1$ \cite{Watanabe}.
$\mathcal{O}_{\rm D}$ should vanish over the line $J_1=J_2$ for an obvious reason
but otherwise non-zero. $\mathcal{O}_{\rm D}$ is found to increase steadily
in AFM $J_2$ region and finally gets saturated in the FM region
at the isotropic point, $\Delta=1$.
On the other hand, it decreases continuously in the anisotropic
region towards the lower values of $\Delta$. Variations of $E_{\rm Gap}$  with  $J_2/J_1$
obtained by exact diagonalization and bond operator formalism are shown in
figure~\ref{spingap}. Exact diagonalization results show that $E_{\rm Gap}$ vanishes
over the line $J_1=J_2$. Once again, bond operator formalism fails to estimate
the value of $E_{\rm Gap}$ close to the line $J_1=J_2$.
It always predicts a non-zero value of $E_{\rm Gap}$
over this line for any value of $\Delta$. In addition, this formalism
underestimates (overestimates) the value of $E_{\rm Gap}$ in FM (AFM) $J_2$ region.

In order to characterize the Haldane phase in $S=1$ Heisenberg chain,
string correlation functions $\mathcal O^\alpha_{\rm S}(i-j)$ and
string order parameter $\mathcal O^\alpha_{\rm S}$ have been
introduced by
den Nijs and Rommelse \cite{Rommelse} and Tasaki~\cite{Tasaki}
and those are defined as
\bea
&&\mathcal O^\alpha_{\rm S}(i-j)=
-\langle S_i^\alpha \,\re^{\ri\piup(S_{i+1}^\alpha +S_{i+2}^\alpha+\ldots+S_{j-1}^\alpha)}\, S_j^\alpha \rangle,\nonumber\\
&&\mathcal O^\alpha_{\rm S}=\lim_{|i-j|\rightarrow \infty}\mathcal O^\alpha_{\rm S}(i-j),
\qquad {\rm where}\qquad \alpha=x,y,z.\nonumber
\eea
Here, $S_i^\alpha$ is the $\alpha$-component of the spin operator $S_i$
with the unity magnitude at the $i$-th site. The $S=\frac{1}{2}$
bond alternating Heisenberg chain can be mapped onto the
isotropic AFM $S=1$ Heisenberg chain when $J_2\rightarrow -\infty$
and $\Delta=1$ \cite{Hida}. Hida also pointed out that bond alternating
Hamiltonian with
anisotropic ($\Delta \neq 1$) $J_1$ bond and isotropic ($\Delta = 1$)
$J_2$ bond can be mapped onto the anisotropic
AFM $S=1$ Heisenberg chain when $J_2\rightarrow -\infty$ \cite{Hidajpsj93}.
In the same way, string correlation functions can
be expressed in terms of $S=\frac{1}{2}$ operators as \cite{Hidaprb92}
\bea
\mathcal O^\alpha_{\rm S}(i-j)=
-4\langle S_{2i}^\alpha \,\re^{\ri\piup(S_{2i+1}^\alpha +S_{2i+2}^\alpha+\ldots+S_{2j-2}^\alpha)} \,S_{2j-1}^\alpha
 \rangle,\qquad \alpha=x,y,z.
\eea
In our model, $\mathcal O^x_{\rm S}(i-j)=\mathcal O^y_{\rm S}(i-j)$ due to the
U(1) symmetry of the Hamiltonian, equation~(\ref{ham}). Values of
$\mathcal O^\alpha_{\rm S}(m)$ for $m=1,2,3,\ldots, 8$
have been estimated numerically on a chain length of $N=32$.
$\mathcal O^\alpha_{\rm S}$ has been obtained by using the
VBS algorithm for extrapolation out of these
$\mathcal O^\alpha_{\rm S}(m)$ values.
For $\Delta = 0$, $\mathcal O^\alpha_{\rm S}$ are symmetric about
$J_2=0$, although this symmetry is lost for $\Delta \neq 0$.
The value of  $\mathcal O^z_{\rm S}$ agrees with the previous estimation
for $\Delta = 1$ \cite{Hida}.  $\mathcal O^z_{\rm S}=\mathcal O^x_{\rm S}$ when $\Delta=1$.
Variations of $\mathcal O^z_{\rm S}$ and $\mathcal O^x_{\rm S}$ have been shown
in figure~\ref{ground_order}~(c) and~(d), respectively. A qualitative similarity
is found in their behaviours even in the anisotropic region.
Both  $\mathcal O^z_{\rm S}$ and $\mathcal O^x_{\rm S}$ are found to decrease
rapidly when $J_2/J_1$ approaches 1.0, and ultimately
vanish exactly over the line  $J_2/J_1=1$. Coexistence of
dimer order and string orders are found throughout the anisotropic region
in this bond alternating Heisenberg chain barring the point $J_2/J_1=-1,\;\Delta=0$.
The spin gap along with the string orders are found to vanish
at the point, $J_2/J_1=-1$, $\Delta=0$, although the dimer order does not.
So, it establishes the fact that the Haldane phase not only exists in bond alternating
Heisenberg chain at the isotropic point, $J_2/J_1\neq 1$, $\Delta=1$
as predicted by Hida \cite{Hida}
but also in most of the anisotropic regions, $J_2/J_1\neq 1$, $0\leqslant\Delta<1$.
In addition, the only point in the anisotropic region where the Haldane phase does not survive is
$J_2/J_1=-1$, $\Delta=0$.
Therefore, apart from the FM point $J_2/J_1=-1$, $\Delta=0$ and AFM line
 $J_2/J_1=1$, $0\leqslant\Delta\leqslant 1$, the Haldane phase exists in the whole parameter regime.
It would be worth mentioning that for FM $J_2$ and $0<\Delta\leqslant 1$,
all the parameters, such as spin gap, dimer and string orders decrease
with an increase of $|J_2/J_1|$ beyond the value $J_2/J_1=-1$.
The nature of decay of those parameters (figure~\ref{ground_order})
indicates that
they all will vanish at larger values of $|J_2/J_1|$ in FM $J_2$ region.
Therefore, this result hints at the collapse of Haldane phase for
larger values of $|J_2/J_1|$ in the full anisotropic region.
Thus, it is expected that either N\'eel or
chiral ordered phase may appear
in the region $|J_2/J_1| \gg 1$, and $0\leqslant \Delta\leqslant1$,
by replacing the Haldane phase.
However, this case is not considered in this study.
\section{Conclusions}
\label{conclusion}
In this work, ground state properties, dispersion relations and spin gap of a
bond alternating anisotropic $S=\frac{1}{2}$ Heisenberg chain
have been evaluated for both the AFM-FM and AFM-AFM cases and in the full
anisotropic regime $0\leqslant \Delta \leqslant 1$.
Both analytic (bond operator and Jordan-Wigner formulations) and numerical
methods are employed to study those properties.
Bond operator and Jordan-Wigner formulations provide more accurate
results in two different parameter regimes. Ground state energy, dispersion relations,
dimer order and spin gap have been derived by bond operator formalism. Longitudinal and
transverse modes of dispersion relations are found.
Longitudinal mode is found to survive even in the absence of
longitudinal part in the Hamiltonian.
For $\Delta=0$, the
exact value of ground state energy has been derived by
using the Jordan-Wigner representation.
Meanwhile, for $\Delta \neq 0$, the ground state energy has been derived by
using the Jordan-Wigner based mean-field theory.
Those theoretical
values have further been supplemented by the
exact diagonalization results and compared to the exact data at
extreme points.
Numerical analysis shows that the
ground state is non-degenerate ($S_{\rm T}^z =0$), while the
first excited state is doubly degenerate ($S_{\rm T}^z =\pm 1$)
for both the cases and throughout the regime $0\leqslant \Delta < 1$.
Although the ground state remains unique, spin gap is found to develop
in the excitation spectrum as soon as the non-uniformity
is introduced in AFM-AFM chain. The spin gap remains non-zero
in most of the AFM-FM region. The non-uniformity of bond strengths
in a bond alternating system breaks
the full translational symmetry of the
model. The gap attributes to the breaking of this translational symmetry
which ultimately gives rise to the Haldane phase.
Spin gap, string orders and dimer order have been obtained numerically.
Spin gap and string orders are found to coexist and non-zero throughout the parameter regime
apart from the point $J_2/J_1=-1$, $\Delta=0$ and line
 $J_2/J_1=1$, $0<\Delta<1$. This phenomenon attributes to the existence of Haldane phase.
Thus, the Haldane phase is present in the whole parameter regime apart
from the point $J_2/J_1=-1$, $\Delta=0$ and line
 $J_2/J_1=1$, $0\leqslant\Delta\leqslant 1$, like the existence of the same at the
isotropic point, $J_2/J_1\neq 1$, $\Delta=1$.
In other words, the Haldane phase is not
only present at the isotropic point but in most of the anisotropic
regime of the bond alternating spin-1/2 Heisenberg chain.
However, the nature of decay of the parameters $E_{\rm Gap}$, $\mathcal O_{\rm D}$,
$\mathcal O^z_{\rm S}$ and $\mathcal O^x_{\rm S}$ indicates that
they all will vanish at larger values of $|J_2/J_1|$ beyond
$J_2/J_1=-1$ in FM $J_2$ region
for $0< \Delta\leqslant1$, which hints at the collapse of Haldane phase.
For this case, it is expected that either N\'eel or
chiral ordered phase may appear
in that region by replacing the Haldane phase.

\subsection{Acknowledgements}
AKG acknowledges the BRNS-sanctioned
research project, 37(3)/14/16/2015, India.

\ukrainianpart
\title{Властивості основного стану спін-$\frac{1}{2}$ анізотропного гайзенбергівського ланцюжка з перемінними зв'язками%
}
\author{С. Пол\refaddr{label1},
       А.К. Гош\refaddr{label2}}
\addresses{
\addr{label1} Фізичний факультет, Шотландський церковний коледж, Колката
 700006, Індія
\addr{label2} Фізичний факультет, Джадавпурський університет,
 Колката 700032, Індія
}

\makeukrtitle

\begin{abstract}
Досліджено властивості основного стану, дисперсійні співвідношення і скейлінгову поведінку спінової щілини спін-$\frac{1}{2}$ анізотропного гайзенбергівського ланцюжка з перемінними зв'язками, коли обмінна взаємодія на навперемінних зв'язках є феромагнітною (FM)
і антиферомагнітною (AFM) в двох окремих випадках.
Результуючі моделі порізно представляють ланцюжки з сусідніми (NN) AFM-AFM і AFM-FM навперемінними зв'язками.
Енергію основного стану оцінено аналітично за допомогою представлення  оператора зв'язку так і представлення Джордана-Вігнера, а також чисельно, використовуючи точну діагоналізацію.
Отримано дисперсійні співвідношення, спінову щілину і декілька  типів впорядкування основного стану. Знайдено, що димерне впорядкування і стрічкові впорядкування співіснують в основному стані.
Знайдено, що спінова щілина появляється як тільки вводиться неоднорідність сили навперемінних  зв'язків в AFM-AFM ланцюжку, яка далі залишається ненульовою для  AFM-FM ланцюжка.
Ця спінова щілина вздовж стрічкових впорядкувань є характерною ознакою фази Галдейна. Знайдено, що фаза Галдейна існує в більшості анізотропної області подібно до ізотропної точки.

\keywords навперемінні зв'язки, спінова щілина, оператор зв'язку, стрічкові впорядкування, димерне впорядкування, скейлінговий закон
\end{abstract}


\begin{thebibliography}{99}
\bibitem{Griffiths}  Griffiths~R.B., Phys. Rev., 1964, {\bf 133},  A768,
\bibdoi{10.1103/PhysRev.133.A768}.
\bibitem{Haldane}  Haldane~F.D.M., Phys. Rev. Lett., 1983, {\bf 50}, 1153,
\bibdoi{10.1103/PhysRevLett.50.1153}.
\bibitem{Rommelse} Den Nijs~M., Rommelse~K., Phys. Rev. B, 1989, {\bf 40}, 4709,
\bibdoi{10.1103/PhysRevB.40.4709}.
\bibitem{Tasaki}Tasaki~H., Phys. Rev. Lett., 1991, {\bf 66}, 798,
\bibdoi{10.1103/PhysRevLett.66.798}.
\bibitem{LSM}Lieb~E.,  Schultz~T., Mattis~D.,
Ann. Phys., 1961, {\bf 16}, 407, \bibdoi{10.1016/0003-4916(61)90115-4}.
\bibitem{LSMA}Affleck~I., Lieb~E.H., Lett. Math. Phys., 1986, {\bf  12}, 57, \bibdoi{10.1007/BF00400304}.
\bibitem{Hastings} Hastings~M.B., Phys. Rev. B, 2004, {\bf 69}, 104431,
\bibdoi{10.1103/PhysRevB.69.104431}.
\bibitem{Bulaevskii}Bulaevskii~L.N., Sov. Phys. JETP, 1963, {\bf 17}, 684.
\bibitem{Harris}Brooks Harris~A., Phys. Rev. B, 1973, {\bf 7}, 3166,
\bibdoi{10.1103/PhysRevB.7.3166}.
\bibitem{Southern}Southern~B.W., Mart\'inez Cu\'ellar J.L., Lavis D.A., Phys. Rev. B, 1998, {\bf 58}, 9156,
\bibdoi{10.1103/PhysRevB.58.9156}.
\bibitem{Kohmoto}Kohmoto~M., Tasaki~H.,  Phys. Rev. B, 1992, {\bf 46}, 3486,
\bibdoi{10.1103/PhysRevB.46.3486}.
\bibitem{Totsuka}Totsuka~K., Phys. Lett. A, 1997, {\bf 228}, 103,
\bibdoi{10.1016/S0375-9601(97)00087-X}.
\bibitem{AKLT1} Affleck~I., Kennedy T., Lieb E.H., Tasaki H., Phys. Rev. Lett., 1987, {\bf 59}, 799,
\bibdoi{10.1103/PhysRevLett.59.799}.
\bibitem{AKLT2} Affleck~I., Kennedy T., Lieb E.H., Tasaki H., Commun. Math. Phys., 1988, {\bf 115}, 477, \bibdoi{10.1007/BF01218021}.
\bibitem{Hida}Hida~K., Phys. Rev. B, 1992, {\bf 45}, 2207,
\bibdoi{10.1103/PhysRevB.45.2207}.
\bibitem{Sakai}Sakai~T., J. Phys. Soc. Jpn., 1995, {\bf 64}, 251,
\bibdoi{10.1143/JPSJ.64.251}.
\bibitem{Hase}Hase~M., Terasaki I., Uchinokura K., Phys. Rev. Lett., 1993, {\bf 70}, 3651,
\bibdoi{10.1103/PhysRevLett.70.3651}.
\bibitem{Jecobs}Jacobs~I.S., Bray J.W., Hart H.R. (Jr.), Interrante L.V., Kasper J.S., Watkins G.D., Prober D.E., Bonner J.C., Phys. Rev. B, 1976, {\bf 14}, 3036,
\bibdoi{10.1103/PhysRevB.14.3036}.
\bibitem{Cross}Cross~M.C., Fisher~D.S., Phys. Rev. B, 1979, {\bf 19}, 402,
\bibdoi{10.1103/PhysRevB.19.402}.
\bibitem{Huizinga}Huizinga~S., Kommandeur J., Sawatzky G.A., Thole B.T., Kopinga K., de Jonge W.J.M., Roos J., Phys. Rev. B, 1979, {\bf 19}, 4723,
\bibdoi{10.1103/PhysRevB.19.4723}.
\bibitem{Yamaguchi}Yamaguchi~H., Shinpuku Y., Shimokawa T., Iwase K., Ono T., Kono Y., Kittaka S., Sakakibara T., Hosokoshi~Y., Preprint \arxiv{1502.06804v1}, 2015.
\bibitem{Watanabe}Watanabe~S., Yokoyama~H., J. Phys. Soc. Jpn., 1999, {\bf 68}, 2073,
\bibdoi{10.1143/JPSJ.68.2073}.
\bibitem{Kodama}Kodama~K., Harashina H., Sasaki H., Kato M., Sato M., Kakurai K., Nishi M., J. Phys. Soc. Jpn., 1999, {\bf 68}, 237,
\bibdoi{10.1143/JPSJ.68.237}.
\bibitem{Miura}Miura~Y., Hirai R., Kobayashi Y., Sato M., J. Phys. Soc. Jpn., 2006, {\bf 75}, 084707,
\bibdoi{10.1143/JPSJ.75.084707}.
\bibitem{Manaka}Manaka~H., Yamada I., Honda Z., Katori H.A., Katsumata K., J. Phys. Soc. Jpn., 1998, {\bf 67}, 3913,
\bibdoi{10.1143/JPSJ.67.3913}.
\bibitem{Stone}Stone~M.B., Tian W., Lumsden M.D., Granroth G.E., Mandrus D., Chung J.-H., Harrison N., Nagler S.E., Phys. Rev. Lett., 2007, {\bf 99}, 087204,
\bibdoi{10.1103/PhysRevLett.99.087204}.
\bibitem{White}White~S.R., Affleck~I., Phys. Rev. B, 1996, {\bf 54}, 9862,
\bibdoi{10.1103/PhysRevB.54.9862}.
\bibitem{sachdev}Sachdev~S., Bhatt~R.N., Phys. Rev. B, 1990, {\bf 41}, 9323,
\bibdoi{10.1103/PhysRevB.41.9323}.
\bibitem{J-W}Jordan~P., Wigner~E., Z. Phys., 1928, {\bf 47}, 631, \bibdoi{10.1007/BF01331938}.
\bibitem{Verkholyak}Verkholyak~T., Honecker~A., Brenig~W.,
Eur. Phys. J. B, 2006, {\bf 49}, 283,
\bibdoi{10.1140/epjb/e2006-00077-1}.
\bibitem{Wang}Wang~Y.R., Phys. Rev. B, 1992, {\bf 46}, 151,
\bibdoi{10.1103/PhysRevB.46.151}.
\bibitem{Lanczos}Grosso~G., Martinelli~L., Parravicini G.P., Phys. Rev. B, 1995, {\bf 51}, 13033,
\bibdoi{10.1103/PhysRevB.51.13033}.
\bibitem{VBS} Broeck~J.-M.V., Schwartz~L.W., SIAM J. Math. Anal., 1979, {\bf 10}, 658, \bibdoi{10.1137/0510061}.
\bibitem{BST}Bulirsch~R., Stoer~J., Numer. Math., 1964, {\bf 6}, 413, \bibdoi{10.1007/BF01386092}.
\bibitem{Hidajpsj93}Hida~K., J. Phys. Soc. Jpn., 1993, {\bf 62}, 1463,
\bibdoi{10.1143/JPSJ.62.1463}.
\bibitem{Hidaprb92}Hida~K., Phys. Rev. B, 1992, {\bf 46}, 8268,
\bibdoi{10.1103/PhysRevB.46.8268}.

\end{thebibliography}
\end{document}